# Mining Educational Data to Analyze Students' Performance

Brijesh Kumar Baradwaj
Research Scholor,
Singhaniya University,
Rajasthan, India

Saurabh Pal
Sr. Lecturer, Dept. of MCA,
VBS Purvanchal University,
Jaunpur-222001, India

*Abstract*— **The main objective of higher education institutions is to provide quality education to its students. One way to achieve highest level of quality in higher education system is by discovering knowledge for prediction regarding enrolment of students in a particular course, alienation of traditional classroom teaching model, detection of unfair means used in online examination, detection of abnormal values in the result sheets of the students, prediction about students' performance and so on. The knowledge is hidden among the educational data set and it is extractable through data mining techniques. Present paper is designed to justify the capabilities of data mining techniques in context of higher education by offering a data mining model for higher education system in the university. In this research, the classification task is used to evaluate student's performance and as there are many approaches that are used for data classification, the decision tree method is used here.**

**By this task we extract knowledge that describes students' performance in end semester examination. It helps earlier in identifying the dropouts and students who need special attention and allow the teacher to provide appropriate advising/counseling.**

## I. INTRODUCTION

The advent of information technology in various fields has lead the large volumes of data storage in various formats like records, files, documents, images, sound, videos, scientific data and many new data formats. The data collected from different applications require proper method of extracting knowledge from large repositories for better decision making. Knowledge discovery in databases (KDD), often called data mining, aims at the discovery of useful information from large collections of data [1]. The main functions of data mining are applying various methods and algorithms in order to discover and extract patterns of stored data [2]. Data mining and knowledge discovery applications have got a rich focus due to its significance in decision making and it has become an essential component in various organizations. Data mining techniques have been introduced into new fields of Statistics, Databases, Machine Learning, Pattern Reorganization, Artificial Intelligence and Computation capabilities etc.

There are increasing research interests in using data mining in education. This new emerging field, called Educational Data Mining, concerns with developing methods that discover knowledge from data originating from educational environments [3]. Educational Data Mining uses many techniques such as Decision Trees, Neural Networks, Naïve Bayes, K- Nearest neighbor, and many others.

Using these techniques many kinds of knowledge can be discovered such as association rules, classifications and clustering. The discovered knowledge can be used for prediction regarding enrolment of students in a particular course, alienation of traditional classroom teaching model, detection of unfair means used in online examination, detection of abnormal values in the result sheets of the students, prediction about students' performance and so on.

The main objective of this paper is to use data mining methodologies to study students' performance in the courses. Data mining provides many tasks that could be used to study the student performance. In this research, the classification task is used to evaluate student's performance and as there are many approaches that are used for data classification, the decision tree method is used here. Information's like Attendance, Class test, Seminar and Assignment marks were collected from the student's management system, to predict the performance at the end of the semester. This paper investigates the accuracy of Decision tree techniques for predicting student performance.

## II. DATA MINING DEFINITION AND TECHNIQUES

Data mining, also popularly known as Knowledge Discovery in Database, refers to extracting or "mining" knowledge from large amounts of data. Data mining techniques are used to operate on large volumes of data to discover hidden patterns and relationships helpful in decision making. While data mining and knowledge discovery in database are frequently treated as synonyms, data mining is actually part of the knowledge discovery process. The sequences of steps identified in extracting knowledge from data are shown in Figure 1.

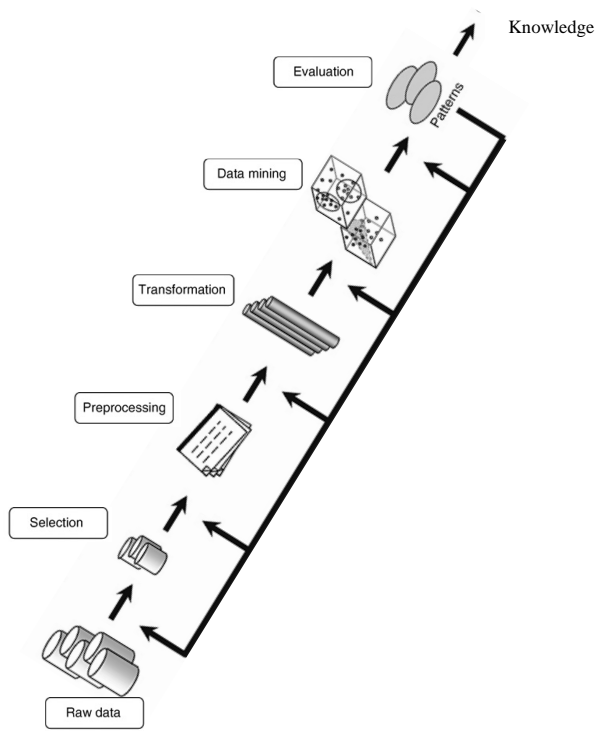

Figure 1: The steps of extracting knowledge from data

Various algorithms and techniques like Classification, Clustering, Regression, Artificial Intelligence, Neural Networks, Association Rules, Decision Trees, Genetic Algorithm, Nearest Neighbor method etc., are used for knowledge discovery from databases. These techniques and methods in data mining need brief mention to have better understanding.

### A. Classification

Classification is the most commonly applied data mining technique, which employs a set of pre-classified examples to develop a model that can classify the population of records at large. This approach frequently employs decision tree or neural network-based classification algorithms. The data classification process involves learning and classification. In Learning the training data are analyzed by classification algorithm. In classification test data are used to estimate the accuracy of the classification rules. If the accuracy is acceptable the rules can be applied to the new data tuples. The classifier-training algorithm uses these pre-classified examples to determine the set of parameters required for proper discrimination. The algorithm then encodes these parameters into a model called a classifier.

### B. Clustering

Clustering can be said as identification of similar classes of objects. By using clustering techniques we can further identify dense and sparse regions in object space and can discover overall distribution pattern and correlations among data attributes. Classification approach can also be used for effective means of distinguishing groups or classes of object but it becomes costly so clustering can be used as preprocessing approach for attribute subset selection and classification.

### C. Predication

Regression technique can be adapted for predication. Regression analysis can be used to model the relationship between one or more independent variables and dependent variables. In data mining independent variables are attributes already known and response variables are what we want to predict. Unfortunately, many real-world problems are not simply prediction. Therefore, more complex techniques (e.g., logistic regression, decision trees, or neural nets) may be necessary to forecast future values. The same model types can often be used for both regression and classification. For example, the CART (Classification and Regression Trees) decision tree algorithm can be used to build both classification trees (to classify categorical response variables) and regression trees (to forecast continuous response variables). Neural networks too can create both classification and regression models.

### D. Association rule

Association and correlation is usually to find frequent item set findings among large data sets. This type of finding helps businesses to make certain decisions, such as catalogue design, cross marketing and customer shopping behavior analysis. Association Rule algorithms need to be able to generate rules with confidence values less than one. However the number of possible Association Rules for a given dataset is generally very large and a high proportion of the rules are usually of little (if any) value.

### E. Neural networks

Neural network is a set of connected input/output units and each connection has a weight present with it. During the learning phase, network learns by adjusting weights so as to be able to predict the correct class labels of the input tuples. Neural networks have the remarkable ability to derive meaning from complicated or imprecise data and can be used to extract patterns and detect trends that are too complex to be noticed by either humans or other computer techniques. These are well suited for continuous valued inputs and outputs. Neural networks are best at identifying patterns or trends in data and well suited for prediction or forecasting needs.

### F. Decision Trees

Decision tree is tree-shaped structures that represent sets of decisions. These decisions generate rules for the classification of a dataset. Specific decision tree methods include Classification and Regression Trees (CART) and Chi Square Automatic Interaction Detection (CHAID).

### G. Nearest Neighbor Method

A technique that classifies each record in a dataset based on a combination of the classes of the k record(s) most similar to it in a historical dataset (where k is greater than or equal to 1). Sometimes called the k-nearest neighbor technique.

### III. RELATED WORK

Data mining in higher education is a recent research field and this area of research is gaining popularity because of its potentials to educational institutes.

Data Mining can be used in educational field to enhance our understanding of learning process to focus on identifying, extracting and evaluating variables related to the learning process of students as described by Alaa el-Halees [4]. Mining in educational environment is called Educational Data Mining.

Han and Kamber [3] describes data mining software that allow the users to analyze data from different dimensions, categorize it and summarize the relationships which are identified during the mining process.

Pandey and Pal [5] conducted study on the student performance based by selecting 600 students from different colleges of Dr. R. M. L. Awadh University, Faizabad, India. By means of Bayes Classification on category, language and background qualification, it was found that whether new comer students will performer or not.

Hijazi and Naqvi [6] conducted as study on the student performance by selecting a sample of 300 students (225 males, 75 females) from a group of colleges affiliated to Punjab university of Pakistan. The hypothesis that was stated as "Student's attitude towards attendance in class, hours spent in study on daily basis after college, students' family income, students' mother's age and mother's education are significantly related with student performance" was framed. By means of simple linear regression analysis, it was found that the factors like mother's education and student's family income were highly correlated with the student academic performance.

Khan [7] conducted a performance study on 400 students comprising 200 boys and 200 girls selected from the senior secondary school of Aligarh Muslim University, Aligarh, India with a main objective to establish the prognostic value of different measures of cognition, personality and demographic variables for success at higher secondary level in science stream. The selection was based on cluster sampling technique in which the entire population of interest was divided into groups, or clusters, and a random sample of these clusters was selected for further analyses. It was found that girls with high socio-economic status had relatively higher academic achievement in science stream and boys with low socio-economic status had relatively higher academic achievement in general.

Galit [8] gave a case study that use students data to analyze their learning behavior to predict the results and to warn students at risk before their final exams.

Al-Radaideh, et al [9] applied a decision tree model to predict the final grade of students who studied the C++ course in Yarmouk University, Jordan in the year 2005. Three different classification methods namely ID3, C4.5, and the NaïveBayes were used. The outcome of their results indicated that Decision Tree model had better prediction than other models.

Pandey and Pal [10] conducted study on the student performance based by selecting 60 students from a degree college of Dr. R. M. L. Awadh University, Faizabad, India. By means of association rule they find the interestingness of student in opting class teaching language.

Ayesha, Mustafa, Sattar and Khan [11] describes the use of k-means clustering algorithm to predict student's learning activities. The information generated after the implementation of data mining technique may be helpful for instructor as well as for students.

Bray [12], in his study on private tutoring and its implications, observed that the percentage of students receiving private tutoring in India was relatively higher than in Malaysia, Singapore, Japan, China and Sri Lanka. It was also observed that there was an enhancement of academic performance with the intensity of private tutoring and this variation of intensity of private tutoring depends on the collective factor namely socio-economic conditions.

Bhardwaj and Pal [13] conducted study on the student performance based by selecting 300 students from 5 different degree college conducting BCA (Bachelor of Computer Application) course of Dr. R. M. L. Awadh University, Faizabad, India. By means of Bayesian classification method on 17 attribute, it was found that the factors like students' grade in senior secondary exam, living location, medium of teaching, mother's qualification, students other habit, family annual income and student's family status were highly correlated with the student academic performance.

### IV. DATA MINING PROCESS

In present day's educational system, a students' performance is determined by the internal assessment and end semester examination. The internal assessment is carried out by the teacher based upon students' performance in educational activities such as class test, seminar, assignments, general proficiency, attendance and lab work. The end semester examination is one that is scored by the student in semester examination. Each student has to get minimum marks to pass a semester in internal as well as end semester examination.

#### A. Data Preparations

The data set used in this study was obtained from VBS Purvanchal University, Jaunpur (Uttar Pradesh) on the sampling method of computer Applications department of course MCA (Master of Computer Applications) from session 2007 to 2010. Initially size of the data is 50. In this step data stored in different tables was joined in a single table after joining process errors were removed.

#### B. Data selection and transformation

In this step only those fields were selected which were required for data mining. A few derived variables were selected. While some of the information for the variables was extracted from the database. All the predictor and response variables which were derived from the database are given in Table I for reference.

TABLE I.        STUDENT RELATED VARIABLES

| Variable | Description | Possible Values |
|---|---|---|
| PSM | Previous Semester Marks | {First > 60% <br> Second >45 & <60% <br> Third >36 & <45% <br> Fail < 36% } |
| CTG | Class Test Grade | {Poor , Average, Good} |
| SEM | Seminar Performance | {Poor , Average, Good} |
| ASS | Assignment | {Yes, No} |
| GP | General Proficiency | {Yes, No} |
| ATT | Attendance | {Poor , Average, Good} |
| LW | Lab Work | {Yes, No} |
| ESM | End Semester Marks | {First > 60% <br> Second >45 & <60% <br> Third >36 & <45% <br> Fail < 36% } |

The domain values for some of the variables were defined for the present investigation as follows:

- **PSM –** Previous Semester Marks/Grade obtained in MCA course. It is split into five class values: *First – >60%, Second – >45% and <60%, Third – >36% and < 45%, Fail < 40%.*

- **CTG –** Class test grade obtained.  Here in each semester two class tests are conducted and average of two class test are used to calculate sessional marks. CTG is split into three classes: *Poor – < 40%, Average – > 40% and < 60%, Good –>60%.*

- **SEM –** Seminar Performance obtained. In each semester seminar are organized to check the performance of students. Seminar performance is evaluated into three classes: *Poor – Presentation and communication skill is low, Average – Either presentation is fine or Communication skill is fine, Good – Both presentation and Communication skill is fine.*

- **ASS –** Assignment performance. In each semester two assignments are given to students by each teacher. Assignment performance is divided into two classes: *Yes – student submitted assignment, No – Student not submitted assignment.*

- **GP -** General Proficiency performance. Like seminar, in each semester general proficiency tests are organized. General Proficiency test is divided into two classes: *Yes – student participated in general proficiency, No – Student not participated in general proficiency.*

- **ATT –** Attendance of Student. Minimum 70% attendance is compulsory to participate in End Semester Examination. But even through in special cases low attendance students also participate in End Semester Examination on genuine reason. Attendance is divided

into three classes: *Poor - <60%, Average - > 60% and <80%, Good - >80%.*

- **LW –** Lab Work. Lab work is divided into two classes: *Yes – student completed lab work, No – student not completed lab work.*

- **ESM -** End semester Marks obtained in MCA semester and it is declared as response variable. It is split into five class values: First – >60% , Second – >45% and <60%, Third – >36% and < 45%, Fail < 40%.

### C. Decision Tree

A decision tree is a tree in which each branch node represents a choice between a number of alternatives, and each leaf node represents a decision.

Decision tree are commonly used for gaining information for the purpose of decision -making. Decision tree starts with a root node on which it is for users to take actions. From this node, users split each node recursively according to decision tree learning algorithm. The final result is a decision tree in which each branch represents a possible scenario of decision and its outcome.

The three widely used decision tree learning algorithms are: ID3, ASSISTANT and C4.5.

### D. The ID3 Decision Tree

ID3 is a simple decision tree learning algorithm developed by Ross Quinlan [14]. The basic idea of ID3 algorithm is to construct the decision tree by employing a top-down, greedy search through the given sets to test each attribute at every tree node. In order to select the attribute that is most useful for classifying a given sets, we introduce a metric - information gain.

To find an optimal way to classify a learning set, what we need to do is to minimize the questions asked (i.e. minimizing the depth of the tree). Thus, we need some function which can measure which questions provide the most balanced splitting. The information gain metric is such a function.

### E. Measuring Impurity

Given a data table that contains attributes and class of the attributes, we can measure homogeneity (or heterogeneity) of the table based on the classes. We say a table is pure or homogenous if it contains only a single class. If a data table contains several classes, then we say that the table is impure or heterogeneous. There are several indices to measure degree of impurity quantitatively. Most well known indices to measure degree of impurity are entropy, gini index, and classification error.

$$\text{Entropy} = \sum_j - p_j \log_2 p_j$$

Entropy of a pure table (consist of single class) is zero because the probability is 1 and log (1) = 0. Entropy reaches maximum value when all classes in the table have equal probability.

$$\text{Gini Index} = 1 - \sum_j p_j^2$$

Gini index of a pure table consist of single class is zero because the probability is 1 and $1\text{-}1^2 = 0$. Similar to Entropy, Gini index also reaches maximum value when all classes in the table have equal probability.

$$\text{Classification Error} = 1 - \max\{p_j\}$$

Similar to Entropy and Gini Index, Classification error index of a pure table (consist of single class) is zero because the probability is 1 and 1-max (1) = 0. The value of classification error index is always between 0 and 1. In fact the maximum Gini index for a given number of classes is always equal to the maximum of classification error index because for a number of classes n, we set probability is equal to $p = \dfrac{1}{n}$ and maximum Gini index happens at $1 - n\dfrac{1}{n^2} = 1 - \dfrac{1}{n}$, while maximum classification error index also happens at $1 - \max\left\{\dfrac{1}{n}\right\} = 1 - \dfrac{1}{n}$.

*F. Splitting Criteria*

To determine the best attribute for a particular node in the tree we use the measure called Information Gain. The information gain, Gain (S, A) of an attribute A, relative to a collection of examples S, is defined as

$$Gain(S, A) = Entropy(S) - \sum_{v \in Values(A)} \frac{|S_v|}{|S|} Entropy(S_v)$$

Where *Values (A)* is the set of all possible values for attribute *A*, and $S_v$ is the subset of *S* for which attribute *A* has value *v* (i.e., $S_v = \{s \in S \mid A(s) = v\}$). The first term in the equation for *Gain* is just the entropy of the original collection *S* and the second term is the expected value of the entropy after S is partitioned using attribute *A*. The expected entropy described by this second term is simply the sum of the entropies of each subset , weighted by the fraction of examples $\dfrac{|S_v|}{|S|}$ that belong to *Gain (S, A)* is therefore the expected reduction in entropy caused by knowing the value of attribute *A*.

$$Split\ Information\ (S, A) = -\sum_{i=1}^{n} \frac{|S_i|}{|S|} \log_2 \frac{|S_i|}{|S|}$$

*and*

$$Gain\ Ratio(S, A) = \frac{Gain(S, A)}{Split\ Information(S, A)}$$

The process of selecting a new attribute and partitioning the training examples is now repeated for each non terminal descendant node. Attributes that have been incorporated higher in the tree are excluded, so that any given attribute can appear at most once along any path through the tree. This process continues for each new leaf node until either of two conditions is met:

1. Every attribute has already been included along this path through the tree, or
2. The training examples associated with this leaf node all have the same target attribute value (i.e., their entropy is zero).

*G. The ID3Algoritm*

ID3 (Examples, Target_Attribute, Attributes)

- Create a root node for the tree
- If all examples are positive, Return the single-node tree Root, with label = +.
- If all examples are negative, Return the single-node tree Root, with label = -.
- If number of predicting attributes is empty, then Return the single node tree Root, with label = most common value of the target attribute in the examples.
- Otherwise Begin
  - o A = The Attribute that best classifies examples.
  - o Decision Tree attribute for Root = A.
  - o For each possible value, $v_i$, of A,
    - ▪ Add a new tree branch below Root, corresponding to the test A = $v_i$.
    - ▪ Let Examples($v_i$) be the subset of examples that have the value $v_i$ for A
    - ▪ If Examples($v_i$) is empty
      - ▪ Then below this new branch add a leaf node with label = most common target value in the examples
    - ▪ Else below this new branch add the subtree ID3 (Examples($v_i$), Target_Attribute, Attributes – {A})
- End
- Return Root

## V. RESULTS AND DISCUSSION

The data set of 50 students used in this study was obtained from VBS Purvanchal University, Jaunpur (Uttar Pradesh) Computer Applications department of course MCA (Master of Computer Applications) from session 2007 to 2010.

TABLE II.    DATA SET

| S. No. | PSM | CTG | SEM | ASS | GP | ATT | LW | ESM |
|--------|------|---------|---------|-----|-----|---------|-----|-------|
| 1. | First | Good | Good | Yes | Yes | Good | Yes | First |
| 2. | First | Good | Average | Yes | No | Good | Yes | First |
| 3. | First | Good | Average | No | No | Average | No | First |
| 4. | First | Average | Good | No | No | Good | Yes | First |
| 5. | First | Average | Average | No | Yes | Good | Yes | First |
| 6. | First | Poor | Average | No | No | Average | Yes | First |

| 7. | First | Poor | Average | No | No | Poor | Yes | Second |
|----|-------|------|---------|-----|-----|------|-----|--------|
| 8. | First | Average | Poor | Yes | Yes | Average | No | First |
| 9. | First | Poor | Poor | No | No | Poor | No | Third |
| 10. | First | Average | Average | Yes | Yes | Good | No | First |
| 11. | Second | Good | Good | Yes | Yes | Good | Yes | First |
| 12. | Second | Good | Good | Yes | Yes | Average | Yes | First |
| 13. | Second | Good | Average | Yes | No | Good | Yes | First |
| 14. | Second | Average | Good | Yes | Yes | Good | No | First |
| 15. | Second | Good | Average | Yes | Yes | Average | Yes | First |
| 16. | Second | Good | Average | Yes | Yes | Poor | Yes | Second |
| 17. | Second | Average | Average | Yes | Yes | Good | Yes | Second |
| 18. | Second | Average | Average | Yes | Yes | Poor | Yes | Second |
| 19. | Second | Poor | Average | No | Yes | Good | Yes | Second |
| 20. | Second | Average | Poor | Yes | No | Average | Yes | Second |
| 21. | Second | Poor | Average | No | Yes | Poor | No | Third |
| 22. | Second | Poor | Poor | Yes | Yes | Average | Yes | Third |
| 23. | Second | Poor | Poor | No | No | Average | Yes | Third |
| 24. | Second | Poor | Poor | No | Yes | Good | Yes | Second |
| 25. | Second | Poor | Poor | Yes | Yes | Poor | Yes | Third |
| 26. | Second | Poor | Poor | No | No | Poor | Yes | Fail |
| 27. | Third | Good | Good | Yes | Yes | Good | Yes | First |
| 28. | Third | Average | Good | Yes | Yes | Good | Yes | Second |
| 29. | Third | Good | Average | Yes | Yes | Good | No | Second |
| 30. | Third | Good | Good | Yes | Yes | Average | Yes | Second |
| 31. | Third | Good | Good | No | Yes | Good | Yes | Second |
| 32. | Third | Average | Average | Yes | Yes | Good | Yes | Second |
| 33. | Third | Average | Average | Yes | No | Average | Yes | Third |
| 34. | Third | Average | Good | No | No | Good | Yes | Third |
| 35. | Third | Good | Average | No | Yes | Average | Yes | Third |
| 36. | Third | Average | Poor | No | No | Average | Yes | Third |
| 37. | Third | Poor | Average | Yes | Yes | Average | Yes | Third |
| 38. | Third | Poor | Average | No | Yes | Poor | Yes | Fail |
| 39. | Third | Average | Average | No | Yes | Poor | Yes | Third |
| 40. | Third | Poor | Poor | Yes | Yes | Good | No | Third |
| 41. | Third | Poor | Poor | No | No | Poor | Yes | Fail |
| 42. | Third | Poor | Poor | No | No | Poor | No | Fail |
| 43. | Fail | Good | Good | Yes | Yes | Good | Yes | Second |
| 44. | Fail | Good | Good | Yes | Yes | Average | Yes | Second |
| 45. | Fail | Average | Good | Yes | Yes | Average | No | Third |
| 46. | Fail | Poor | Good | Yes | Yes | Average | No | Fail |
| 47. | Fail | Good | Good | Yes | Yes | Poor | Yes | Fail |
| 48. | Fail | Poor | Poor | No | No | Poor | Yes | Fail |
| 49. | Fail | Average | Average | Yes | Yes | Good | Yes | Second |
| 50. | Fail | Poor | Good | No | No | Poor | No | Fail |

To work out the information gain for A relative to S, we first need to calculate the entropy of S. Here S is a set of 50 examples are 14 "*First*", 15 "*Second*", 13 "*Third*" and 8 "*Fail*"..

$$
\begin{aligned}
\text{Entropy (S)} =\ & - p_{First} \log_2(p_{First}) - p_{Second} \log_2(p_{Second}) \\
& - p_{third} \log_2(p_{third}) - p_{Fail} \log_2(p_{Fail}) \\
=\ & -\left(\frac{14}{50}\right)\log_2\left(\frac{14}{50}\right) - \left(\frac{15}{50}\right)\log_2\left(\frac{15}{50}\right) \\
& -\left(\frac{13}{50}\right)\log_2\left(\frac{13}{50}\right) - \left(\frac{8}{50}\right)\log_2\left(\frac{8}{50}\right) \\
=\ & 1.964
\end{aligned}
$$

To determine the best attribute for a particular node in the tree we use the measure called Information Gain. The information gain, Gain (S, A) of an attribute A, relative to a collection of examples S,

$$
Gain(S, PSM) = Entropy(S) - \frac{|S_{First}|}{|S|} Entropy(S_{First})
$$

$$
- \frac{|S_{Second}|}{|S|} Entropy(S_{Second}) - \frac{|S_{Third}|}{|S|} Entropy(S_{Third})
$$

$$
- \frac{|S_{Fail}|}{|S|} Entropy(S_{Fail})
$$

TABLE III.　　GAIN VALUES

| Gain | Value |
|------|-------|
| Gain(S, PSM) | 0.577036 |
| Gain(S, CTG) | 0.515173 |
| Gain(S, SEM) | 0.365881 |
| Gain(S, ASS) | 0.218628 |
| Gain (S, GP) | 0.043936 |
| Gain(S, ATT) | 0.451942 |
| Gain(S, LW) | 0.453513 |

PSM has the highest gain, therefore it is used as the root node as shown in figure 2.

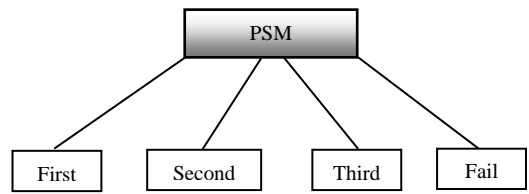

Figure 2. PSM as root node

Gain Ratio can be used for attribute selection, before calculating Gain ratio Split Information is shown in table IV.

TABLE IV.　　SPLIT INFORMATION

| Split Information | Value |
|-------------------|-------|
| Split(S, PSM) | 1.386579 |
| Split (S, CTG) | 1.448442 |
| Split (S, SEM) | 1.597734 |
| Split (S, ASS) | 1.744987 |
| Split (S, GP) | 1.91968 |
| Split (S, ATT) | 1.511673 |
| Split (S, LW) | 1.510102 |

Gain Ratio is shown in table V.

TABLE V.　　GAIN RATIO

| Gain Ratio | Value |
|-----------|-------|
| Gain Ratio (S, PSM) | 0.416158 |
| Gain Ratio (S, CTG) | 0.355674 |
| Gain Ratio (S, SEM) | 0.229 |
| Gain Ratio (S, ASS) | 0.125289 |
| Gain Ratio (S, GP) | 0.022887 |
| Gain Ratio (S, ATT) | 0.298968 |
| Gain Ratio (S, LW) | 0.30032 |

This process goes on until all data classified perfectly or run out of attributes. The knowledge represented by decision tree can be extracted and represented in the form of IF-THEN rules.

| |
| --- |
| IF PSM = 'First' AND ATT = 'Good' AND CTG = 'Good' or 'Average' THEN ESM = First |
| IF PSM = 'First' AND CTG = 'Good' AND ATT = "Good' OR 'Average' THEN ESM = 'First' |
| IF PSM = 'Second' AND ATT = 'Good' AND ASS = 'Yes' THEN ESM = 'First' |
| IF PSM = 'Second' AND CTG = 'Average' AND LW = 'Yes' THEN ESM = 'Second' |
| IF PSM = 'Third' AND CTG = 'Good' OR 'Average' AND ATT = "Good' OR 'Average' THEN PSM = 'Second' |
| IF PSM = 'Third' AND ASS = 'No' AND ATT = 'Average' THEN PSM = 'Third' |
| IF PSM = 'Fail' AND CTG = 'Poor' AND ATT = 'Poor' THEN PSM = 'Fail' |

Figure 3. Rule Set generated by Decision Tree

One classification rules can be generated for each path from each terminal node to root node. Pruning technique was executed by removing nodes with less than desired number of objects. IF- THEN rules may be easier to understand is shown in figure 3.

## CONCLUSION

In this paper, the classification task is used on student database to predict the students division on the basis of previous database. As there are many approaches that are used for data classification, the decision tree method is used here. Information's like Attendance, Class test, Seminar and Assignment marks were collected from the student's previous database, to predict the performance at the end of the semester.

This study will help to the students and the teachers to improve the division of the student. This study will also work to identify those students which needed special attention to reduce fail ration and taking appropriate action for the next semester examination.

## AUTHORS PROFILE

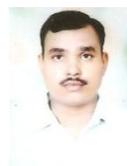

**Brijesh Kumar Bhardwaj** is Assistant Professor in the Department of Computer Applications, Dr. R. M. L. Avadh University Faizabad India. He obtained his M.C.A degree from Dr. R. M. L. Avadh University Faizabad (2003) and M.Phil. in Computer Applications from Vinayaka mission University, Tamilnadu. He is currently doing research in Data Mining and Knowledge Discovery. He has published one international paper.

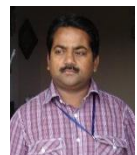

**Saurabh Pal** received his M.Sc. (Computer Science) from Allahabad University, UP, India (1996) and obtained his Ph.D. degree from the Dr. R. M. L. Awadh University, Faizabad (2002). He then joined the Dept. of Computer Applications, VBS Purvanchal University, Jaunpur as Lecturer. At present, he is working as Head and Sr. Lecturer at Department of Computer Applications.

Saurabh Pal has authored a commendable number of research papers in international/national Conference/journals and also guides research scholars in Computer Science/Applications. He is an active member of IACSIT, CSI, Society of Statistics and Computer Applications and working as Reviewer/Editorial Board Member for more than 15 international journals. His research interests include Image Processing, Data Mining, Grid Computing and Artificial Intelligence.